# Resolved-Sideband Raman Cooling to the Ground State of an Optical Lattice.

S. E. Hamann, D. L. Haycock, G. Klose, P. H. Pax, I. H. Deutsch[a] and P. S. Jessen

Optical Sciences Center, University of Arizona, Tucson, AZ 85721

We trap neutral Cs atoms in a two-dimensional optical lattice and cool them close to the zero-point of motion by resolved-sideband Raman cooling. Sideband cooling occurs via transitions between the vibrational manifolds associated with a pair of magnetic sublevels and the required Raman coupling is provided by the lattice potential itself. We obtain mean vibrational excitations $\bar{n}_x \approx \bar{n}_y \approx 0.01$, corresponding to a population ~98% in the vibrational ground state. Atoms in the ground state of an optical lattice provide a new system in which to explore quantum state control and subrecoil laser cooling.

The preparation and coherent manipulation of pure quantum states represents the ultimate control which can be exerted over a physical system. A spectacular example is the preparation of the motional ground state of a trapped atomic ion by resolved sideband cooling [1], which has been followed by the generation of Fock, coherent and squeezed states, and of Schrödinger cat states [2]. More recent developments include the demonstration of quantum logic gates [3], which represents a first step toward quantum computation and the implementation of algorithms as unitary transformations on a quantum system [4]. Various degrees of quantum state control have been achieved also for the electromagnetic field [5], for Rydberg [6] and nuclear [7] wavepackets in atoms and molecules, and in chemical reactions [8]. In this letter we demonstrate for the first time the preparation of the vibrational ground state of optically trapped neutral Cs atoms by resolved-sideband Raman cooling [9]. Of order $10^6$ atoms are individually trapped deep in the Lamb-Dicke regime in independent potential wells of a two-dimensional, far detuned optical lattice [10] formed by the interference of three laser beams, and cooled to a mean vibrational excitation of ~0.01 per degree of freedom. In bringing this degree of quantum control to the atom/optical lattice system, it is our goal to significantly expand its role as a laboratory for the study of non-classical atomic motion and quantum state manipulation.

Our sideband cooling scheme relies on Raman transitions between the vibrational manifolds associated with a pair of magnetic sublevels of the $6S_{1/2}(F=4)$ hyperfine ground state. Compared to ion traps and free space Raman cooling [11], which rely on transitions between different hyperfine ground states, our optical lattice system offers the considerable advantage of using Raman coupling between magnetic sublevels which is intrinsic to the lattice potential itself. This eliminates the need for separate, phase locked Raman lasers separated by frequencies in the GHz regime, and allows for a remarkably simple experimental setup. At the densities used here, atoms in different potential wells do not interact, and it is straightforward to prepare a macroscopic number of atoms in the vibrational ground state. By comparison, the strong Coulomb interaction in ion traps makes it very challenging to cool two or more ions to their common ground state. Because of the efficiency inherent in laser cooling, in excess of 80% of the atoms captured in our magneto-optic trap end up in the motional ground state of the far-off-resonance lattice. This is in dramatic contrast to previous demonstrations of state selection in optical lattices [12]; in those experiments the initial ground state fraction was only a few percent and the selection process therefore very inefficient. Given the combination of simplicity and cooling efficiency, we expect that resolved-sideband Raman cooling in optical lattices, followed by adiabatic expansion [13], will prove an attractive means of optical subrecoil cooling.

The basic design for our optical lattice is a configuration of three coplanar laser beams [14] as shown in Fig. 1a, with equal amplitudes $E_1$ and linear polarizations in the lattice plane. This lattice consists of nearly isotropic potential wells centered at positions where the local polarization is either $\sigma_+$ or $\sigma_-$. The lattice frequency is typically detuned 20 GHz (3831 natural linewidths) below the $6S_{1/2}(F=4) \to 6P_{3/2}(F'=5)$ transition at 852 nm, which is much further than the separation between hyperfine excited states. The optical potential for alkali atoms in the limit of large detuning is discussed in detail in ref. [15], and we give here a brief summary of the salient features. The total electric field in the lattice is $\mathbf{E}_L(\mathbf{x}) = \mathrm{Re}[E_1 \vec{\varepsilon}_L(\mathbf{x}) e^{-i\omega_L t}]$, where $\vec{\varepsilon}_L(\mathbf{x})$ is the local polarization (not necessarily unit norm). The optical potential can then be written in the compact form

$$\hat{U}(\mathbf{x}) = -\frac{2}{3} U_1 |\vec{\varepsilon}_L(\mathbf{x})|^2 \hat{I} + \frac{i}{3} U_1 [\vec{\varepsilon}_L^*(\mathbf{x}) \times \vec{\varepsilon}_L(\mathbf{x})] \cdot \frac{\hat{\mathbf{F}}}{F},$$

where $\hat{I}$, $\hat{\mathbf{F}}$ are the identity and angular momentum operators and $U_1$ is the magnitude of the light shift induced by a single lattice beam driving a transition with unit Clebsch-Gordon coefficient. When the polarizations of all the lattice beams lie entirely in the $x$-$y$ plane, the second term has the form of an effective magnetic field along $\hat{z}$. $\hat{U}(\mathbf{x})$ is then diagonal in the basis of magnetic sublevels, with each diagonal element corresponding to the (diabatic) optical potential for that state. Off-diagonal elements correspond to Raman coupling between magnetic sublevels and can be obtained only by adding to the lattice light field a component polarized along $\hat{z}$ ($\pi$-polarization). Our two-dimensional lattice is the simplest configuration for which this is possible [15]; a $\pi$-polarized component with well defined amplitude $E_\pi$ and phase $e^{i\phi}$ relative to the in-plane component is conveniently introduced by changing the polarization of one of the lattice beams. Considering only Raman coupling between magnetic sublevels $|m=4\rangle$ and $|m=3\rangle$, the operator $\hat{U}(\mathbf{x})$ can be expanded to first order in the Lamb-Dicke parameter $\eta = \sqrt{E_R/\hbar\omega}$ around a potential minimum. With a convenient choice of lattice beam phases that puts a $\sigma_+$-polarized potential well at the origin, and choosing $\varphi = \pi/2$ to maximize the Raman coupling, we find

$$\hat{U}_{4,3}(\mathbf{x}) \approx -i \frac{U_1}{4\sqrt{2}} \frac{E_\pi}{E_1} (2 + kX - i 3kY),$$

where $X$ and $Y$ are the quantum mechanical position operators. From this expression we see that the off-diagonal elements of the potential provide Raman coupling between adjacent vibrational levels, and therefore allows for sideband cooling along both $\hat{x}$ and $\hat{y}$.

In the tight binding regime the bound states of the lattice are well approximated by product states $|n,m\rangle = |n\rangle \otimes |m\rangle$, where $|n\rangle$ is a two-dimensional harmonic oscillator state with total vibrational excitation $n = n_x + n_y$ associated with the diabatic potential for the magnetic sublevel $|m\rangle$. Motion in the direction normal to the lattice plane is separable and can be ignored. In the experiment discussed below, the amplitude of the out-of-plane polarized component is $E_\pi \approx 0.3 E_1$, the single-beam light shift is $U_1 \approx 54 E_R$ (where $E_R = (\hbar k)^2/2m$ is the photon recoil energy), and the Lamb-Dicke parameter is $\eta \approx 0.2$. A typical value for the coupling matrix element $\langle n-1, m=3 | \hat{U}_{4,3}(\mathbf{x}) | n, m=4 \rangle$ is then $\approx 0.7 \sqrt{n_x} E_R$ and $\approx 2 \sqrt{n_y} E_R$ for $\Delta n_x = -1$ and $\Delta n_y = -1$ respectively. For comparison, the maximum light shift is $U_0 \approx 243 E_R$, and the oscillation frequency in the $|m=4\rangle$ potential is $\hbar\omega \approx 20 E_R$.

Resolved-sideband Raman cooling can now be accomplished as illustrated in Fig. 1b. We add a weak magnetic field $B_z$ along $\hat{z}$ to Zeeman shift states $|n, m=4\rangle$ and $|n-1, m=3\rangle$ into degeneracy; in the terminology of sideband cooling, this corresponds to tuning the Raman coupling to the first red sideband. The lattice then stimulates transitions $|n, m=4\rangle \leftrightarrow |n-1, m=3\rangle$. Relaxation from $|n-1, m=3\rangle$ to $|n-1, m=4\rangle$ is provided by optical pumping, which is accomplished with a pair of $\sigma_+$-polarized pumper and repumper beams resonant with the $F=4 \to F'=4$ and $F=3 \to F'=4$ hyperfine transitions. Optical pumping on the $F=4 \to F'=4$ transition serves to decouple atoms in the $|n, m=4\rangle$ states from the resonant pumper light, so that $|n=0, m=4\rangle$ is a dark state. This comes at the price of occasional decay to the $F=3$ hyperfine ground state, which is problematic because the minima for the $F=4$ potentials coincide with saddle points for the $F=3$ potentials. The presence of an intense, resonant repumper beam ensures that the atoms are returned to the $F=4$ state before their wavepackets can disperse, so that no significant heating occurs. Solution of the rate equations for optical pumping shows that, on average, roughly one pumper and one repumper photon is necessary to effect the transition $|n-1, m=3\rangle \to |n-1, m=4\rangle$.

Our sideband cooling experiment proceeds as follows. A vapor cell magneto-optic trap and 3D optical molasses is used to prepare a cold (3 μK) sample of ~$10^6$ atoms in a volume ~400 μm in diameter. After the 3D molasses beams are extinguished, the atoms are cooled and localized in a near-resonance 2D optical lattice with the same beam configuration as the far-off-resonance lattice. The atoms are then adiabatically transferred to the superimposed, far-off-resonance lattice by decreasing the depth of the near-resonance lattice to zero and simultaneously increasing the depth of the far-off-resonance lattice from zero to the final operating depth. This produces a sample of atoms localized in the far-off-resonance lattice, deep in the Lamb-Dicke regime and with minimal vibrational excitation [16]. When the transfer is completed, we begin resolved-sideband Raman cooling by adding a field $B_z$ to tune the lattice Raman coupling to the red sideband and turning on the pumper/repumper beams. After some time, cooling is terminated by a short (~40 μs) interval of intense optical pumping in order to transfer as many atoms as possible to $|m=4\rangle$. We then quickly extinguish the lattice, pumper and repumper beams and measure the atomic momentum distribution along the vertical direction. This is accomplished by a standard Time-Of-Flight (TOF) analysis, which infers the momentum distribution from the atomic arrival times at a ~50 μm thick probe beam located 4.7 cm below the lattice (Fig. 1a). The observed momentum distributions are generally indistinguishable from a Gaussian fit, and we characterize them by the corresponding kinetic temperature $T = \langle p^2 \rangle / k_B M$. Rotating the lattice with respect to gravity allows us to obtain momentum distributions in different directions. In practice we are constrained to measure the momentum distribution along one of the lattice beams, i. e. along $\hat{y}$ and the direction $\hat{x}'$ forming a 30° angle with the $x$-axis. Because the atoms occupy states that are close to harmonic, this is sufficient to determine that they are two-dimensionally cold. Adding a magnetic field gradient during the measurement separates the TOF distributions for different states $|m\rangle$ and allows us to determine the distribution of population over magnetic sublevels [17].

When they have first been loaded into the far-off-resonance lattice, atoms are distributed fairly evenly among all magnetic sublevels. A measurement after 11 ms of sideband cooling shows, however, that at least 90% of the atoms have been optically pumped into the stretched state $|m=4\rangle$ (Fig. 2b). In the figure, apparent residual population in states $|m \leq 3\rangle$ is an artifact caused by atoms escaping from the 3D molasses and optical lattice prior to the TOF measurement, and by adiabatic transitions between magnetic sublevels as the atoms are released from the lattice. Because we finish the cooling sequence with a short, intense optical pumping pulse we expect that the final population in $|m=4\rangle$ is in fact very close to 100%. It is then straightforward to obtain a Boltzmann factor for the vibrational populations from the measured kinetic temperature. First we calculate the kinetic temperature $T_0$ of the vibrational ground state in the $|m=4\rangle$ potential (including a correction for anharmonicity), based on a very careful estimate of the lattice light intensity and detuning. For our parameters ($U_1 = 54(3) E_R$) we find $T_0 = 951(30)$nK [18]. The potential wells are sufficiently isotropic that there is no significant variation in $T_0$ along different directions. Using a harmonic model the Boltzmann factor is then given by $q_B = (T - T_0)/(T + T_0)$, where the effect of anharmonicity is taken into account via the correction in $T_0$. Fig. 2a shows a typical momentum distribution measured after 11 ms of sideband cooling. For comparison we also show the calculated ground state distribution, as well as an uncooled distribution. It is immediately apparent that the atoms are cooled very close to the zero-point of motion. After the sideband cooling process has reached steady state, there is no measurable difference between the momentum spread along $\hat{x}'$ and $\hat{y}$, nor is there any measurable difference in the steady state temperature at detunings of 10 GHz, 20 GHz and 35 GHz; the data discussed below was all measured for a detuning of 20 GHz and in the $\hat{x}'$ direction. There is however a strong dependence on the repumper frequency, which must be tuned to the $F = 3 \to F' = 4$ to better than a natural linewidth. This is readily explained based on the need to repump atoms before they disperse in the $F = 3$ potential. To determine the minimum temperature that can be reached in our experiment we performed a series of 16 measurements similar to Fig. 2a in the course of roughly one hour. From this data we obtain an average kinetic temperature $T = 0.966(10)\,\text{nK}$ [18]. The corresponding Boltzmann factor is $q_B = 0.008(16)$, and the mean vibrational excitation per degree of freedom is $\bar{n}_x \approx \bar{n}_y \approx 0.008(16)$. For atoms in $|m = 4\rangle$ this corresponds to a population $\pi_0 = 0.984(31)$ in the two-dimensional vibrational ground state.

Further information about the sideband cooling can be gained from a measurement of the steady state vibrational excitation versus Raman detuning. The insert of Fig. 3 shows the kinetic temperature as a function of the tuning field $B_z$, and clearly demonstrates cooling on both the first and second red sideband. In steady state we can invoke detailed balance to obtain $1/q_B = \pi_0/\pi_1 = \Gamma_{cool}/\Gamma_{heat}$. The heating rate $\Gamma_{heat}$ is roughly independent of $B_z$, since it is dominated by extraneous processes such as spontaneous scattering of lattice photons, while the cooling rate $\Gamma_{cool}$ is proportional to the rate of Raman absorption on the red sideband. We therefore expect Lorentzian maxima in $1/q_B$ when $B_z$ corresponds to Raman resonance, which is indeed what we see in Fig. 3. The separation between the center of the two cooling resonances is equal to the energy difference between the first and second excited vibrational manifolds of $|m = 4\rangle$, and so provides an independent measurement of the lattice depth and of the ground state kinetic temperature $T_0$. On this basis we find $T_0 = 997(50)\,\text{nK}$, which agrees nicely with the value based on our estimate of the lattice light intensity. The full-width at half maximum of the Lorentzian fit to the first sideband is $\sim 3 E_R$, within 40% of the estimated width of the $|n = 1, m = 3\rangle$ state caused by optical pumping. This agreement is reasonable considering the uncertainty on $1/q_B$ around the peak of the resonance.

In conclusion we have demonstrated resolved-sideband Raman cooling to the ground state of a two-dimensional far-off-resonance optical lattice. Our method relies on Raman coupling intrinsic to the lattice potential and uses a magnetic field to tune the coupling to the "red sideband" as required for cooling. This results in a simple setup which yields ground state populations around 98%. Extension of the scheme to three-dimensional lattices is straightforward and should result in comparable vibrational temperatures. A wide range of experiments on non-classical motion will benefit from the initial preparation of atoms in a pure quantum state, including squeezing, collapse and revival of wavepackets [19], the study of quantum chaotic motion [20], and the study of quantum transport and quantum tunneling in optical potentials [21]. Conditions in the optical lattice are also favorable for quantum state manipulation, and we are currently working on experiments to prepare both Fock states and states that are coherently distributed over more than one potential well. Finally, neutral atoms near the electronic ground state interact very weakly with each other, and the techniques explored here could be applied in large-periodicity lattices where each potential well contains several atoms. Cooling of a few atoms to a common ground state would provide a new route to quantum degeneracy, in a regime very different from Bose-Einstein condensation [22].

This work was supported by NSF Grant No. PHY-9503259, by ARO Grant No. DAAG559710165, and by JSOP Grant No. DAAG559710116.

**Figure Captions**

**FIG. 1**. (a) Experimental setup for resolved-sideband Raman cooling in a two-dimensional optical lattice. The basic lattice is formed by three coplanar laser beams with linear polarizations in the lattice plane. To add a $\pi$-component to the lattice light field and provide Raman coupling we change the polarization of one beams to elliptical. The pumper and repumper beams are $\sigma_+$ polarized, normal to the lattice plane and counterpropagating, in order to balance radiation pressure. (b) Basic cooling scheme. The far-off-resonance lattice potential induces $\sigma_\pm \leftrightarrow \pi$ stimulated Raman transitions and couple the magnetic sublevels $|m=3,4\rangle$ of the $F=4$ hyperfine ground state. Relaxation $|m=3\rangle \rightarrow |m=4\rangle$ is provided by a pair of pumper and repumper beams. Resolved-sideband Raman cooling accumulates most of the atoms in the vibrational ground state of the stretched state $|m=4\rangle$.

**FIG. 2**. (a) Momentum distribution in the far-off-resonance lattice, measured in the $x'$ direction. Solid line: following 11 ms of resolved-sideband Raman cooling. Open circles: calculated distribution for the vibrational ground state. Dashed line: atoms trapped for 11 ms without cooling. (b) Stern-Gerlach analysis of the atomic internal state. Top curve: magnetic population following 11 ms of resolved-sideband Raman cooling. The stretched state contains ~90% of the population, with most of the remainder in $|m=3\rangle$. Bottom curve: magnetic populations before cooling.

**FIG. 3**. Inverse Boltzmann factor as a function of applied magnetic field (Raman detuning). Cooling on the first and second sideband is evident. Solid circles are data points, the solid line a fit to the sum of two Lorentzians. Insert: corresponding kinetic temperatures. The dashed line indicates the kinetic temperature of the vibrational ground state, the solid line is a transformation of the fit from the main figure.

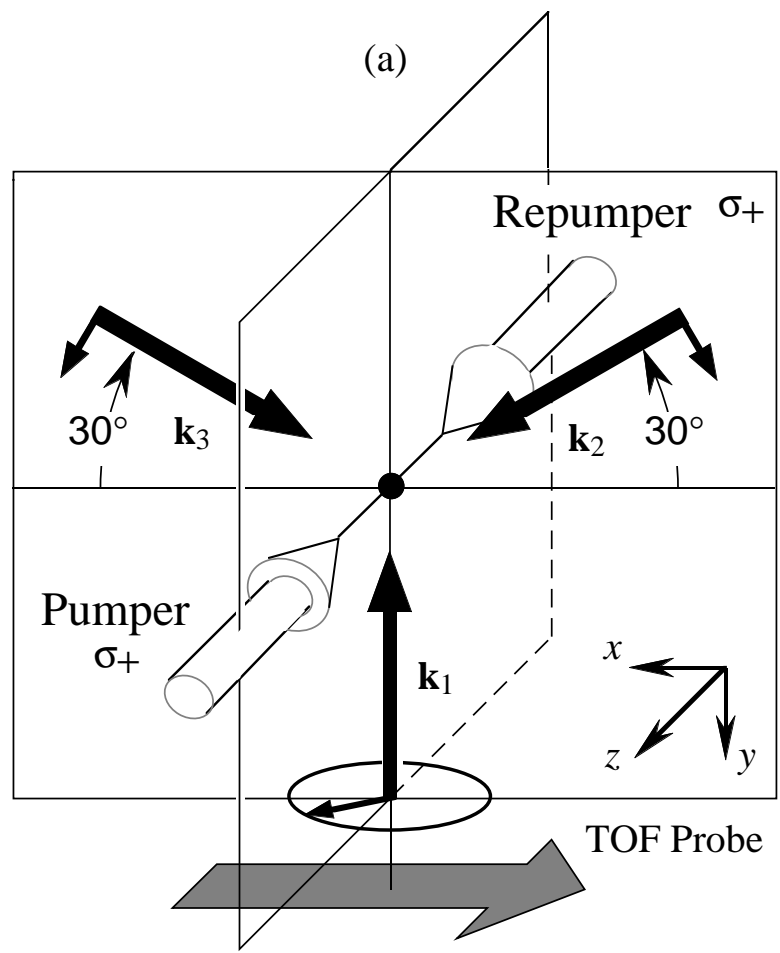
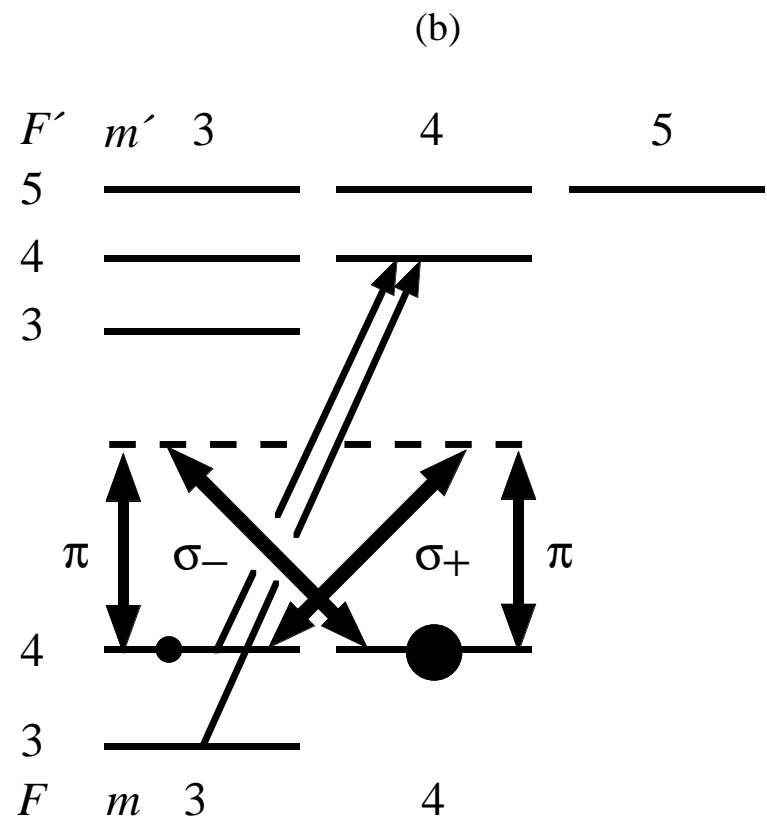

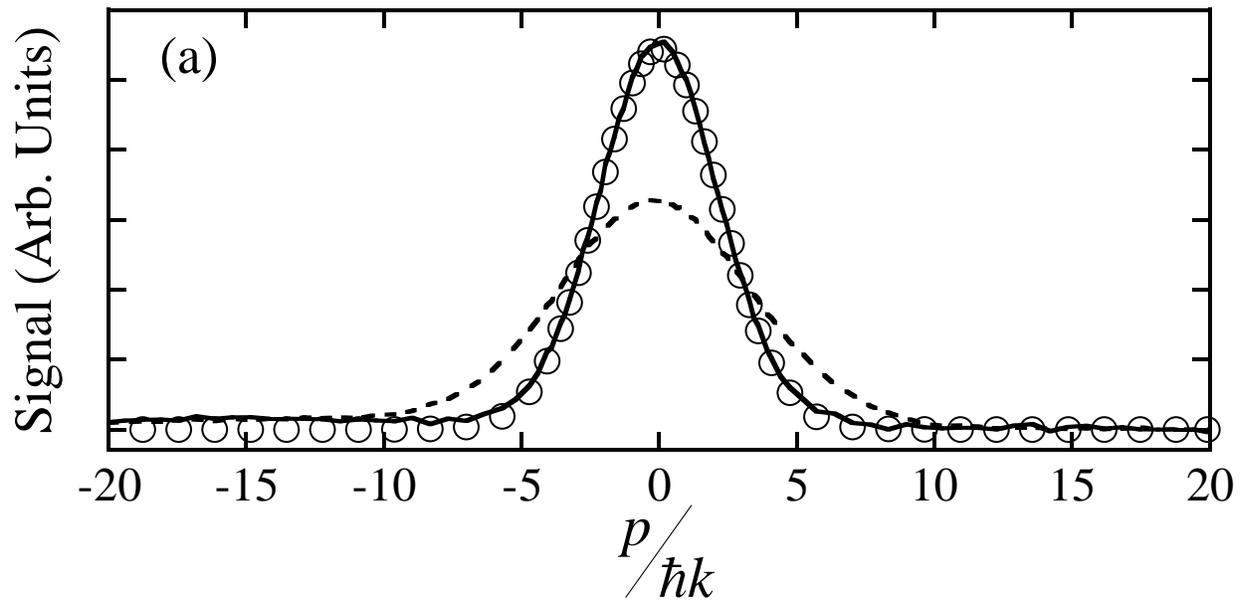
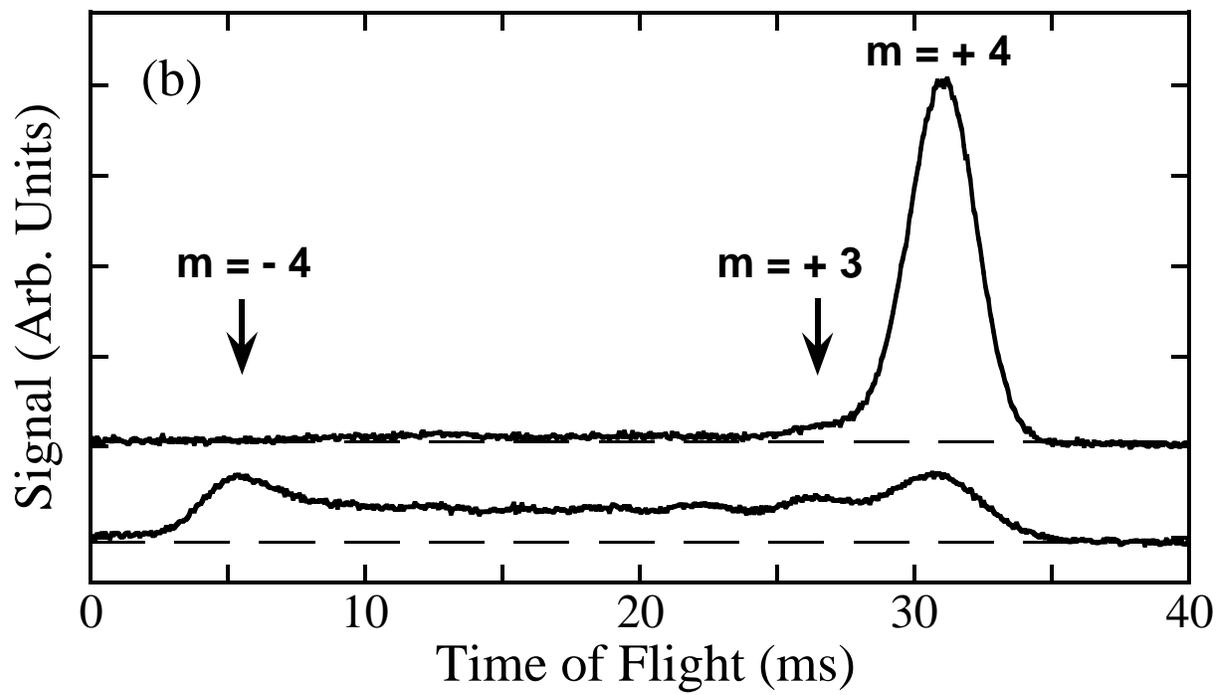

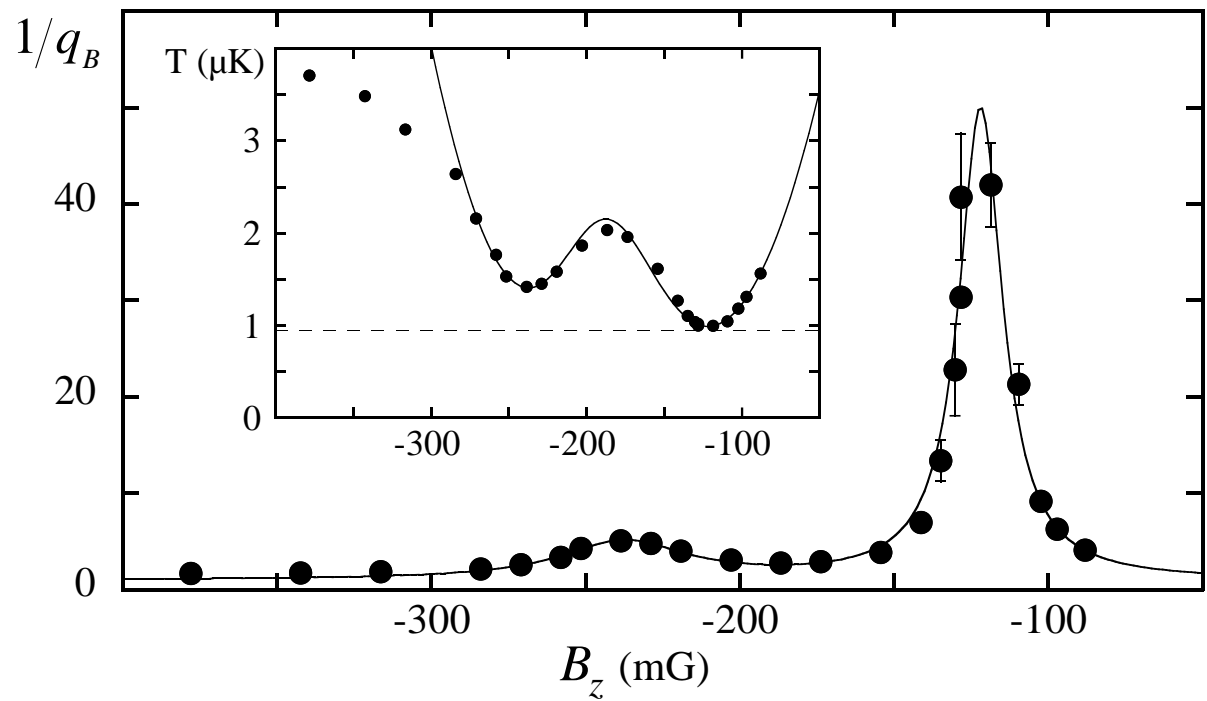